\documentclass{aip-cp}
\usepackage[numbers]{natbib}
\usepackage{rotating}
\usepackage{graphicx}
\usepackage{epstopdf}
\usepackage{ulem}
\usepackage{amssymb}
\usepackage{times}
\usepackage{graphicx,bm,amssymb}

\def\aj{AJ}
\def\apj{ApJ}
\def\apjl{ApJ}
\def\apjs{ApJS}
\def\aap{A\&A}
\def\aaps{A\&AS}
\def\mnras{MNRAS}
\def\physrep{Phys.~Rep.}

\newcommand{\be}{\begin{equation}}
\newcommand{\ee}{\end{equation}}
\newcommand{\bse}{\begin{subequations}}
\newcommand{\ese}{\end{subequations}}
\newcommand{\bary}{\begin{eqnarray}}
\newcommand{\eary}{\end{eqnarray}}

\begin{document}

\title{Fermi LAT observation of VER J2019+407}
\author[aff1]{M. Araya\corref{cor1}}
\author[aff2]{N. Fraija}
\eaddress{nifraija@astro.unam.mx}
\affil[aff1]{Escuela de F\'isica \& Centro de Investigaciones Espaciales (CINESPA), Universidad de Costa Rica, San Jos\'e 2060, Costa Rica}
\affil[aff2]{Instituto de Astronom\'ia, Universidad Nacional Aut\'{o}noma de M\'{e}xico, Apdo. Postal 70-264, Cd. Universitaria, DF 04510, M\'{e}xico}
\corresp[cor1]{Corresponding author: miguel.araya@ucr.ac.cr}
\maketitle

\begin{abstract}
Analysis of data from the Fermi Large Area Telescope (LAT) in the region of the supernova remnant G78.2+2.1 reveals an excess at the position of the TeV source VER J2019+407. The GeV source is extended with a hard spectrum and likely the counterpart of the TeV source. The spectrum of the LAT emission connects smoothly with that of VER J2019+407, and the overall GeV--TeV spectrum is best described by a broken power-law with indices 1.8 and 2.5 below and above a break energy of 71 GeV. Several scenarios are considered to explain the nature of this unidentified gamma ray source.
\end{abstract}

\section{INTRODUCTION}
The mechanism of diffusive shock acceleration is thought to operate at the shocks of supernova remnants (SNRs) \cite[e.g.,][]{1978MNRAS.182..147B,1987PhR...154....1B}. There is observational evidence for the presence of high-energy leptons \cite[e.g.,][]{1994AstL...20..157A,2001ApJ...552L..39G,2002A&A...395..943B,2002ApJ...581.1101H,2002ApJ...581.1116R} and hadronic cosmic rays in SNRs \cite[e.g.,][]{2009ApJ...706L...1A,2010ApJ...722.1303A,2010ApJ...718..348A,2010ApJ...710L..92A,2010ApJ...712..459A,2010Sci...327.1103A,2010ApJ...712..287E,2011ApJ...735..120Y,2015ApJ...813....3A,2013Sci...339..807A}, although it is not yet clear whether these sources can reach the highest energies in the Galactic cosmic rays. Many $\gamma$-ray sources such as VER J2019+407 \citep{2013ApJ...770...93A} remain unidentified. With a TeV extent of $0^{\circ}.23\pm0^{\circ}.03_{\mbox{\tiny stat}}\,^{+0^{\circ}.04}_{-0^{\circ}.02\mbox{\tiny sys}}$ and a power law TeV spectrum \citep{2013ApJ...770...93A}, it is located within the SNR G78.2+2.1. Non-thermal radio emission is also seen in the region \citep{1977AJ.....82..329H,1977AJ.....82..718H,1980A&AS...39..133L,2008A&A...490..197L} as well as X-ray emission \citep{2013MNRAS.436..968L,2006A&A...457.1081K,1997A&A...324..641Z,1996MNRAS.281.1033B,1977AJ.....82..329H,1966ApJ...144..937D}.

The SNR has been extensively studied across the electromagnetic spectrum \cite[e.g.,][]{1988A&A...191..313P,2006A&A...457.1081K,2003AJ....125.3145T,2008A&A...490..197L,1980A&AS...39..133L,1986A&AS...63..345B,2001AstL...27..233G,1983AJ.....88...97H,1988A&A...191..313P,2004A&A...427L..21B,2002ApJ...571..866U,2013MNRAS.436..968L,2011A&A...529A.159G}. In $\gamma$-rays, it has also been seen by the EGRET instrument on board the CGRO \citep{1995ApJS..101..259T} and the Large Area Telescope (LAT) on borad Fermi \citep{2012ApJ...756....5L}. A very bright $\gamma$-ray pulsar, PSR J2021+4026, is thougth to be associated with the remnant \citep{2010ApJS..187..460A,2010ApJS..188..405A}. In the latest catalog of LAT sources, the morphology of the SNR is described by a uniform disc of radius 0$^{\circ}$.63 centered at the position (J2000) $\alpha = 305^\circ.27$, $\delta = 40^\circ.52$ \citep{2015ApJS..218...23A}.

The presence of the pulsar in the region makes the analysis of LAT data quite challenging. However, by observing at high enough energies where the pulsar is dim, we are able to identify an additional component of the $\gamma$-ray emission in the region of the SNR G78.2+2.1 which is in spatial conincidence with the TeV source VER J2019+407. We are therefore able to obtain the overall spectral energy distribution of the source across three orders of magnitude in the high and very-high energy regimes.

\section{DATA ANALYSIS AND RESULTS}\label{LATdata}
LAT data in the region are analyzed from 2008 August to 2015 November in the energy interval 4--300 GeV with the publicly available software version v10r0p5 and the instrument response functions (IRFs) P8R2\_SOURCE\_V6P. Standard cuts are applied and the spectral properties are obtained by the method of maximum likelihood \citep{1996ApJ...461..396M}. After including the disc describing the SNR, the residuals show a clear excess at the position of the TeV source. In order to study this emission further, a test statistic (TS) map is obtained above 15 GeV by moving a point source at each pixel position in the predefined map grid and evaluating its TS, without including the cataloged LAT sources representing the disc nor the pulsar PSR J2021+4026 in the model.

\begin{figure}[h]
  \centerline{\includegraphics[width=290pt]{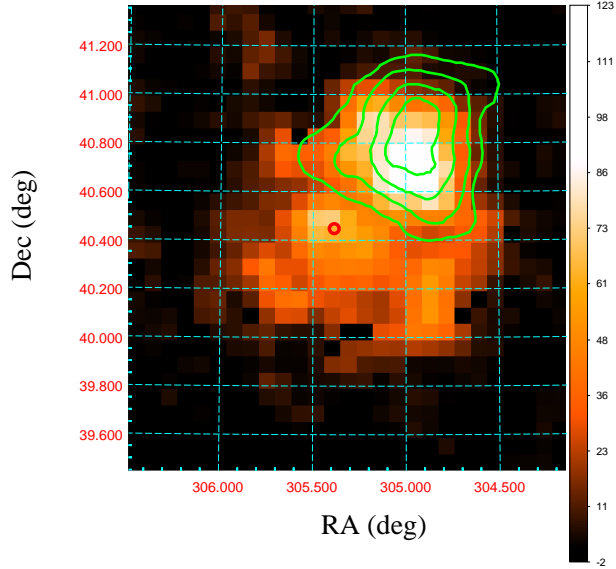}}
\caption{TS map of the region containing the SNR G78.2+2.1 for LAT data above 15 GeV and a $0.067$ deg pixel$^{-1}$ scale. The location of the TeV emission as seen by VERITAS is indicated by the contours and that of the $\gamma$-ray pulsar PSR J2021+4026 with a circle.\label{fig1}}
\end{figure}

The result is shown in Figure \ref{fig1} with the TeV photon excess contours from the VERITAS observation \citep{2013ApJ...770...93A}. The morphology used by the VERITAS Collabotarion to describe the TeV emission is adopted to obtain the spectrum with LAT data (a two-dimensional Gaussian, see \citep{2013ApJ...770...93A}), while the rest of the emission is described with a template consisting of the original disc in the LAT catalog \citep{2015ApJS..218...23A}, 3FGL J2021.0+4031e, with a $0^{\circ}.23$-radius circular region around VER J2019+407 removed.

In the 4--300 GeV range the TS value for VER J2019+407 is 180 and a simple power-law fit yields a spectral index of $1.91\pm0.10_{\mbox{\tiny stat}}$ and a flux of $(7.61\pm0.84_{\mbox{\tiny stat}})\times10^{-10}$ ph cm$^{-2}$ s$^{-1}$. In this energy range the LAT spectrum is not curved since the same TS value is obtained after using a log parabola and a power law with an exponential cutoff for the spectral shape, both of which contain an additional degree of freedom with respect to the simple power law model.

\begin{figure}[h]
  \centerline{\includegraphics[width=400pt]{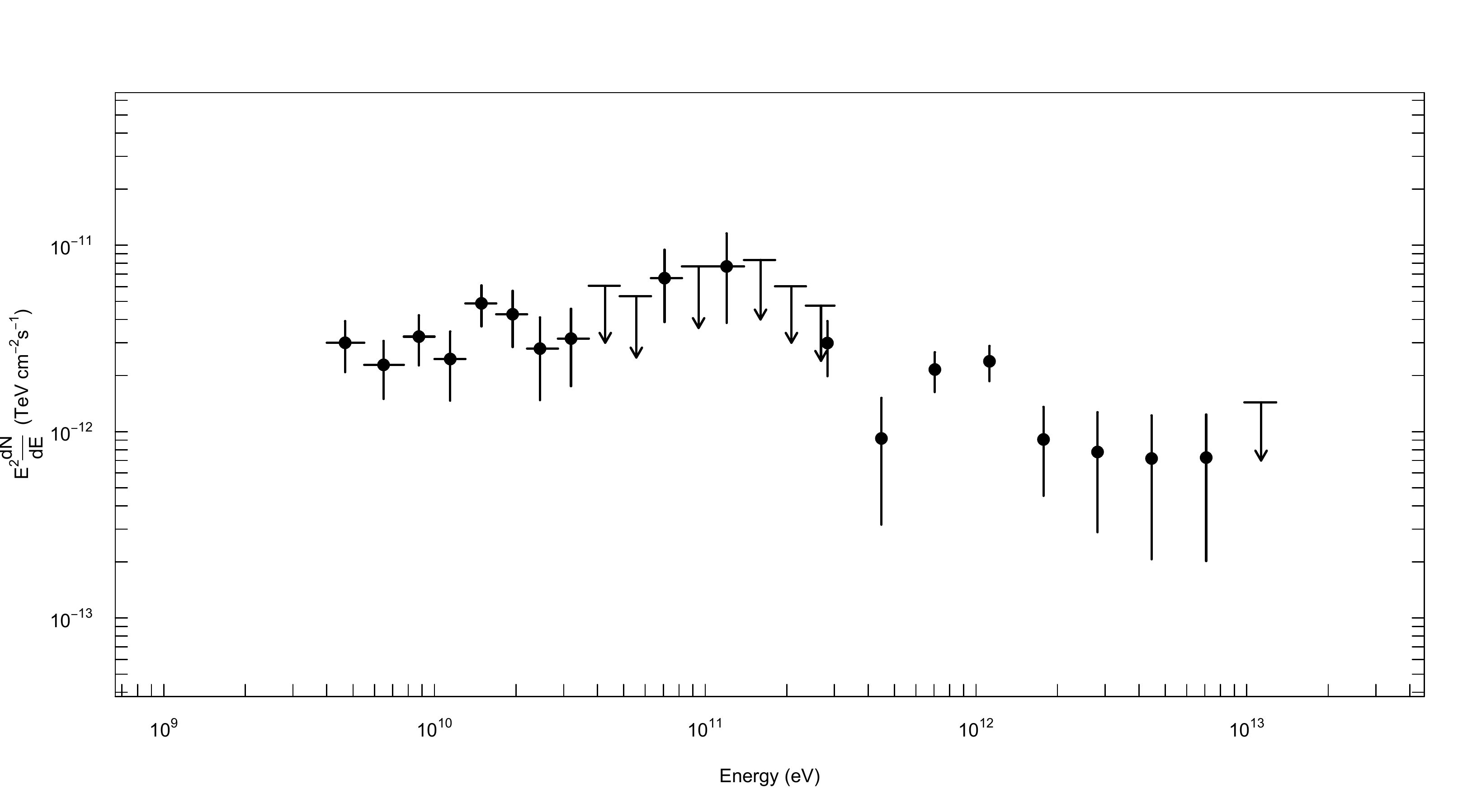}}
\caption{$\gamma$-ray spectral energy distribution of VER J2019+407. The LAT points are obtained in this work and the TeV points from a VERITAS observation.\label{fig2}}
\end{figure}

In order to construct a spectral energy distribution the LAT data is binned in 16 energy intervals and a maximum likelihood fit is performed in each while keeping the spectral index fixed to calculate the source flux. A 95\% confidence level upper limit is calculated in the intervals for which the source TS falls below 9. The result is shown in Figure \ref{fig2} together with the VERITAS data points from \citep{2013ApJ...770...93A}. The LAT spectrum seems to connect smoothly with the VERITAS spectrum, which supports the scenario for a common origin of the $\gamma$-rays.

The effect of freeing the bright pulsar (PSR J2021+4026) spectral parameters is seen to have a $\sim$10\% effect in the lowest LAT energy bin, which is added as a systematic error. The effect of the pulsar in the other bins is found to be much smaller. A detailed analysis of the spectrum of VER J2019+407 below 4 GeV is left for a future work. A preliminary analysis above 1 GeV hints to a possible spectral softening, although we cannot know at this point if this is an artifact caused by emission from the pulsar \citep{2016ApJ...826...31F}.

\section{DISCUSSION}\label{discussion}
The spectrum of VER J2019+407 does not show curvature in the energy range 4--300 GeV but this is not so when coupled with the data at TeV energies, as can be easily seen in Figure \ref{fig2}. We fit the GeV--TeV data and find that a broken power law best describes the photon spectrum. The resulting $\gamma$-ray indices below and above the break are $1.78\pm 0.15$ and $2.46\pm 0.11$, respectively, and the break energy is $70.7\pm 0.1$ GeV (reduced $\chi^2 = 1.02$) \citep{2016ApJ...826...31F}.

In the context of a hadronic scenario, this photon spectrum requires a parent particle population that is also a broken power-law with a relatively hard distribution below the break \citep{2016ApJ...826...31F}. Such a distribution could be accounted for by nonlinear effects and wave damping \citep{1999ApJ...511L..53M, 1999ApJ...526..385B, 2009ApJ...706L...1A, 1999ApJ...513..311B}.

\begin{figure}[h]
\centerline{\includegraphics[width=400pt]{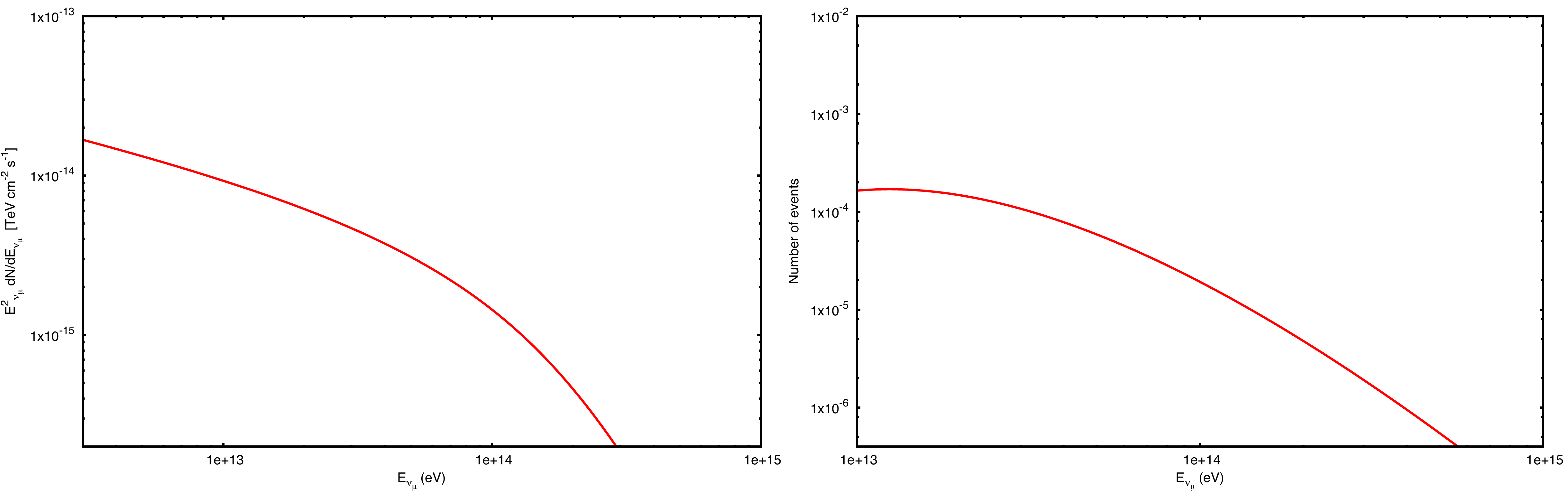}}
\caption{Neutrino spectrum (left panel) and number of events (right panel) as a function of neutrino energy expected in the IceCube experiment from VER J2019+407 in a hadronic scenario for the $\gamma$-rays.\label{fig3}}
\end{figure}

From the cosmic ray population required to explain the $\gamma$-ray data \citep{2016ApJ...826...31F}, we estimated the number of reconstructed neutrino events in the IceCube experiment. The left panel in Figure \ref{fig3} shows the neutrino flux as a function of neutrino energy. The right panel in Figure \ref{fig3} shows the number of neutrinos as a function of neutrino energy expected in the IceCube experiment. The implication is that this detector cannot record events from this source.

A leptonic scenario could also explain the $\gamma$-ray fluxes. It has been already proposed that nonthermal bremsstrahlung emission can explain the hard X-ray fluxes in the region \cite{2002ApJ...571..866U} and could also account for a substantial component of the radiation around a few GeV, while inverse Compton scattering of soft photons by the same high energy electrons could explain the TeV fluxes. The model from \cite{2002ApJ...571..866U} in the bremsstrahlung-dominated scenario clearly overpredicts the observed fluxes in Figure \ref{fig2}, which indicates that the target density must be much lower than their value of $10$ cm$^{-3}$. We estimate that a density of 1 cm$^{-3}$ and the same shape for the electron distribution solves this conflict \citep{2016ApJ...826...31F}.

Finally, we point out that in order to confirm or discard the leptonic scenario, additional observations of the region are required above 11 GHz to quantify the amount of synchrotron emission and thus help constrain the maximum lepton energy.

\section{ACKNOWLEDGMENTS}
We are grateful to PAPIIT-UNAM IG100414 and Universidad de Costa Rica for financial support. This research has made use of NASA's Astrophysical Data System and the Canadian Galactic Plane Survey (CGPS) which is a Canadian project with international partners. The Dominion Radio Astrophysical Observatory is operated as a national facility by the National Research Council of Canada. The CGPS is supported by a grant from the Natural Sciences and Engineering Research Council of Canada.
%
%

\nocite{*}
\bibliographystyle{aipnum-cp}%
\bibliography{arayaCP}%

\end{document}